\documentstyle[prc,aps,epsfig]{revtex}

\begin{document}
\title{
Shape of the 4.438 MeV $\gamma$-ray line of $^{12}$C from proton and
$\alpha$-particle induced reactions on  $^{12}$C and  $^{16}$O}

\draft

\author{J. Kiener, N. de Sereville, V. Tatischeff} 
\address{CSNSM, IN2P3-CNRS et Universit\'e Paris-Sud,
B\^{a}timents 104 et 108,  F-91405 Campus Orsay, France}

\date{\today}

\maketitle

\begin{abstract} We calculated in detail the angular distribution
of $\gamma$-rays and the resulting shape of the $\gamma$-ray line produced
by the nuclear deexcitation of the 4.439 MeV state of $^{12}$C following
proton and $\alpha$-particle interactions with $^{12}$C and $^{16}$O in the
energy range from threshold to 100 MeV per nucleon, making use of available
experimental data. In the proton energy range from 8.6 to 20 MeV, the
extensive data set of a recent accelerator experiment on $\gamma$-ray line
shapes and angular distributions was used to deduce parametrizations for the
$\gamma$-ray emission of the 2$^+$, 4.439 MeV state of $^{12}$C following
inelastic proton scattering off $^{12}$C and proton induced spallation of
$^{16}$O. At higher proton energies and for $\alpha$-particle induced
reactions, optical model calculations were the main source to obtain the
needed reaction parameters for the calculation of $\gamma$-ray line
shapes and angular distributions. Line shapes are predicted for various
interaction scenarios of accelerated protons and $\alpha$-particles in solar
flares.
\end{abstract}


\section{Introduction}

The $\gamma$-ray line at 4.438 MeV from the deexcitation of the first
excited state of $^{12}$C is one of the strongest nuclear lines, clearly
visible in several energetic solar flares, which were observed in particular
by the Gamma Ray Spectrometer aboard the SMM satellite
\cite{Atlas} and one of the best candidates for an observation at
$\gamma$-ray energies of the interaction of low energy cosmic rays in nearby
molecular clouds \cite{CP}. It is mainly produced by inelastic scattering of
energetic protons and $\alpha$-particles off $^{12}$C and by the spallation
of $^{16}$O by the same particles \cite{RKL}. This line is also produced in
reverse kinematics by accelerated $^{12}$C and $^{16}$O bombarding the
ambient hydrogen and helium nuclei, with however a large Doppler broadening
due to the high velocities of the emitting $^{12}$C nucleus, which makes its
detection much more difficult. In the following we will refer to
$\gamma$-ray lines produced by light ion bombardment of the ambient gas as
the narrow component and to $\gamma$-rays produced by accelerated heavy ions
as the broad component.

Accordingly, the intensity and shape of the narrow component of the 4.438
MeV line depend on the properties of the accelerated light particle spectrum
and the $^{12}$C and $^{16}$O abundances in the interaction site. The narrow
line intensities from various isotopes can be used to derive accelerated
particle and ambient gas abundances, as it has been done for example by
Murphy et al. \cite{MRK} for the solar flare of 1981, April 27. More
detailed information on the energy spectrum of the accelerated particles and
their directional distribution at the interaction site may however only be
obtained by a detailed line shape analysis.

Murphy, Kozlovsky and Ramaty \cite{MKR} investigated the impact of different
accelerated particle angular distributions on the narrow component of the
4.438 MeV $\gamma$-ray line produced by the p + $^{12}$C reaction. The shape
of the broad component has been discussed by Bykov, Bozhokin and Bloemen
\cite{BBB} and by Kozlovsky, Ramaty and Lingenfelter \cite{KRL}. These
studies are based on the experimentally well-known excitation functions,
some available experimental data on differential cross sections, few
published results on measured line shapes, and theoretical arguments. For
the spallation reaction, the situation is similar with respect to available
experimental data. The excitation function is experimentally determined from
threshold up to over 100 MeV, but few data are available on differential
cross sections and line shapes.

In order to put the calculations for the 4.438 MeV line on a firmer basis,
we use additionally to the already published data the extensive data set on
$\gamma$-ray angular distributions and $\gamma$-ray line shapes for the
$^{12}$C(p,p$\gamma$) and $^{16}$O(p,p$\alpha \gamma$) reaction, obtained in
a recent accelerator experiment at Orsay \cite{KB}. For the spallation
reaction, the reproduction of the observed line shapes for proton energies
between 16 and 20 MeV proved to be only sensitive to the mean excitation
energy of the $\alpha$-particle emitting intermediate states of $^{16}$O. 
Estimations
of the mean excitation energy for higher proton energies and for the
differential cross sections are based on published data in the proton energy
range 40-50 MeV and optical model (OM) calculations.

For proton inelastic scattering, we adopted a method similar to that of
Murphy, Kozlovsky and Ramaty \cite{MKR} for line shape calculations. Two
independent amplitudes for the magnetic substate population of the 2$^+$
level and a phase shift were adjusted to reproduce both the experimental
line shapes and the laboratory $\gamma$-ray angular distributions in the
proton energy range from 8.6 MeV to 19.75 MeV. The extrapolations to higher
energies are based on OM calculations using phenomenological potential
parameters, which proved to give an excellent reproduction of measured
differential cross sections for nucleon elastic and inelastic scattering off
$^{12}$C at projectile energies E $\approx$ 20 - 100 MeV
\cite{Mei}.

$\alpha$-particle inelastic scattering off $^{12}$C and spallation of
$^{16}$O suffer from a complete absence of published data on line shapes.
Therefore, estimations of the mean excitation energy in $^{16}$O for the
spallation reaction were entirely based on OM calculations and an
experimental $\alpha$-particle spectrum at $E_{\alpha}$ = 60 MeV. For the
inelastic scattering reaction, differential cross sections
d$\sigma$/d$\Omega_{\alpha}$ and some laboratory $\gamma$-ray angular
distributions have been used to find the parameters for the $\gamma$-ray
emission based on the population amplitude method at $\alpha$-particle 
energies below
20 MeV, while at higher energies the needed parameters were obtained from OM
calculations.

In the last section we discuss the production of the 4.438 MeV line in solar
flares and show some calculated shapes for the narrow component of the 4.438
MeV line induced by proton and $\alpha$-particle interactions in the solar
atmosphere as it may be observed by the HESSI spacecraft \cite{Hessi}.

\section{The spallation reaction $^{16}O(\lowercase{p,x}\gamma)^{12}C$ }

Total cross section data are reported by Dyer et al. \cite{Dyer} for $E_p$ =
14-23 MeV, Lesko et al. \cite{Les} for $E_p$ = 20, 30, 33, 40 and 50 MeV,
Lang et al. \cite{LW} at $E_p$ = 40 MeV, Zobel et al. \cite{Zob} at $E_p$ =
12.1, 28.2, 48.3 and 145 MeV and by Foley et al. \cite{Fol} at $E_p$ = 146
MeV. From threshold to 23 MeV, we used the data of Ref. \cite{Dyer}, 
which are in
agreement with those of Refs. \cite{Les} and \cite{Zob}. 
There is, however, some
disagreement between the different data in the energy interval $E_p$ = 40-50
MeV. The cross section values at $E_p$ = 40 MeV of Ref. \cite{LW} and Ref.
\cite{Les} differ by more than a factor of two. The same holds for the
values around 50 MeV of Ref. \cite{Les} and Ref. \cite{Zob}. As no
systematic deviation of one of the data sets can be observed, no cross
section value was discarded. We therefore fitted the excitation function
curve above 23 MeV. Assigning the same weight to each data point, the
overall shape of the excitation function curve from 23 MeV to 146 MeV
could be best fitted by a power law plus a constant. The fit is very close
to the curve in Fig.3 of Ref. \cite{RKL} for this reaction. The obtained cross
sections values are given in Table \ref{tab1}.

For proton energies below 50 MeV, that reaction should proceed principally
from a sequential process with the excitation of intermediate states of
$^{16}$O, decaying subsequently to $\alpha$ + $^{12}$C$^{\star}_{4.439}$.
Reactions with other particles than the $\alpha$-particle in the outgoing
channel, like $^3$He + n or $^3$H + p have thresholds $\approx$ 20 MeV
higher, around $E_p$ = 35 MeV, and should be negligible for the production
of $^{12}$C$^{\star}_{4.439}$ in solar flares. The sequential character can
be deduced from a study of the $^{16}$O(p,p$\alpha$)$^{12}$C reaction at a
proton energy of 46.8 MeV, where no evidence for quasi-elastic p-$\alpha$
scattering was found \cite{EQB}. Still other reaction paths, such as
$^{16}$O(p,$\alpha$)$^{13}$N$^{\star}$ $\rightarrow$ p +
$^{12}$C$^{\star}_{4.439}$ were also found negligible. At lower energies
these reaction mechanisms should be still weaker. We adopted in the
following exclusively the sequential process
$^{16}$O(p,p')$^{16}$O$^{\star}$ $\rightarrow$ $\alpha$ +
$^{12}$C$^{\star}_{4.439}$.

Parameters entering into the line shape determination are therefore the
relative contribution of the different excited states in $^{16}$O and for
each excited state the differential cross section with respect to the
recoiling $^{16}$O$^{\star}$ scattering angle, the
$^{12}$C$^{\star}$-$^{16}$O angular correlation and the $\gamma$-$^{12}$C
angular correlation. Given the scarcity of experimental data and the number
of excited states involved, it is highly impracticable if not impossible to
undergo a full and correct calculation for numerous proton energies from
threshold to 100 MeV. One can, however, reasonably assume that the average
angular distribution of the $\alpha$-$^{12}$C$^{\star}$ decay in the
recoiling $^{16}$O$^{\star}$ system and the angular distribution of the
$\gamma$-ray emission in the $^{12}$C$^{\star}$ system are isotropic,
because of the contribution of many $^{16}$O levels with different spin and
parity (see the discussion in Ref. \cite{LW}).

\subsection{$E_p$ = 16 - 20 MeV}
\vspace{0.2 cm}

In this energy range, experimental line shapes are available from a recent
accelerator experiment at Orsay \cite{KB} at six different $\gamma$-ray
detection angles, ranging from 45$^{\circ}$ to 145$^{\circ}$ with respect to
the incoming beam direction and at four proton energies. These experimental
line shapes were compared with calculated ones obtained with the use of a
Monte-Carlo type program. In this program reaction parameter inputs were the
differential cross section with respect to the proton center-of-mass
scattering angle and the mean excitation energy in $^{16}$O. As mentionned
above, the vector of the $\alpha$-$^{12}$C$^{\star}$ decay in the
$^{16}$O$^{\star}$ system and the $\gamma$-ray emission in the
$^{12}$C$^{\star}$ system were taken isotropically. Slowing down of the
carbon in the target with exponentially decaying $\gamma$-ray emission
probability corresponding to the lifetime of the 4.439 MeV state, as well as
the full detector geometry and resolution were taken into account to
generate theoretical detector spectra. Stopping powers were taken from the
code TRIM \cite{TRIM}.

Optical-model calculations of inelastic proton scattering off $^{16}$O to
states from 12.75 MeV to 17 MeV, with optical potential parameters from the
compilation of Perey and Perey \cite{Per} were performed to obtain the
differential inelastic scattering cross sections
d$\sigma$/d$\Omega_{p_{cm}}$. The calculated cross sections 
were found to be roughly isotropic in this proton energy range. As
furthermore the calculated line shapes proved to be not very sensitive to
the details of the differential cross section, we simply used a constant
d$\sigma$/d$\Omega_{p_{cm}}$. The mean excitation energy $\mathcal{E}$$_x$
in $^{16}$O was then the only parameter that has been adjusted to reproduce
the experimental $\gamma$-ray spectra.

Figure \ref{fig1} shows the result of a line shape calculation at 
$E_p$ = 20 MeV for
the six detection angles of the experiment. Position, line shape and
relative intensities of the experimental lines are fairly well reproduced by
the calculation, which gives credit to the validity of the assumption of
essentially isotropic angular distributions. We also checked that
calculations with a Gaussian distribution of the excitation energy centered
at $\mathcal{E}$$_x$ resulted in practically identical line shapes.  Those
calculations have been made at the four proton energies where experimental
spectra were extracted. The obtained mean excitation energies are presented
in Table \ref{tab1}.

\subsection{$E_p$ = 20-100 MeV}
\vspace{0.2 cm}

In this proton energy range, some experimental information on line shapes
and on inelastic cross sections of excited levels in $^{16}$O are available
at $E_p$ = 40, 45 and 46.8 MeV \cite{LW} \cite{BMS} \cite{EQB}.
Lang et al. \cite{LW} deduced relative contributions to the
$^{12}$C$^{\star}_{4.439}$ production for $^{16}$O states ranging from $E_x$
= 12.53 MeV to 28 MeV from the cross sections at $\Theta_p$ = 40$^{\circ}$
and $E_p$ = 45 MeV of Buenerd et al. \cite{BMS}. For excitation energies
above 28 MeV, virtually no contribution has been observed in the
$^{16}$O(p,p')$^{16}$O reaction at $E_p$ = 45 MeV \cite{BMS} and in the
$^{16}$O(p,p$\alpha$)$^{12}$C$^{\star}_{4.439}$ reaction at $E_p$ = 46.8 MeV
\cite{EQB}.

We used as the basis of our calculations eight representative states $E_i$
for the $^{16}$O levels, with the relative contribution to the
$^{12}$C$^{\star}_{4.439}$ production $P_i$:

\begin{equation}
P_i = \frac{\Gamma^i_{\alpha_1}}{\Gamma^i_{tot}} 
\frac{d\sigma_i}{d\Omega}(40^{\circ})
\end{equation}

given by Lang et al. \cite{LW} in their table I for $E_p$ = 45 MeV. Here,
$\Gamma^i_{\alpha_1}$ and $\Gamma^i_{tot}$ mean the partial decay width to the
$\alpha$ + $^{12}$C$^{\star}_{4.439}$ channel and the total width of the
state $i$, respectively. In detail, we chose the following representative
states with excitation energy $E_i$:

(1) $E_1$ = 12.75 MeV, 2$^+$, representing the 12.53, 13.02 and 13.26 MeV
levels of Ref. \cite{LW}. 

(2) $E_2$ = 15.1 MeV, 2$^+$, representing the 13.97, 15.27 and 15.50
MeV levels of Ref. \cite{LW}

(3) $E_{3-8}$ = 17, 19, 21, 23, 25 and 27 MeV, 2$^+$, representing the
respective excitation energy range $E_x$ = $E_i$ $\pm$ 1 MeV of Ref. \cite{LW}.

The choice of exclusively {\it l}=2 excitations to 2$^+$ states has two
reasons. First, calculations of inelastic proton scattering to the giant
resonance region at $E_x$ = 21-28 MeV indicate good agreement of {\it l}=2
excitation with the experimental data \cite{BMS}. Second, the dependence of
the cross section on the proton energy proved to be not very sensitive to
the multipolarity, and the angular dependence of the cross section for {\it
l}=2 excitations is somehow intermediate between {\it l}=1 and {\it l}=3
excitations, simulating fairly well an averaged angular distribution for the
three multipolarities. Non-natural parity states and excitations with {\it
l}$>$3 are believed not to change much the overall trends.

We also assumed that the total cross section for each state is proportional
to the differential cross section at 40$^{\circ}$. This was supported by OM
calculations for $E_p$ = 45 MeV, where the ratio of the inelastic 
scattering cross section
$\sigma_i$ to the differential cross section $d\sigma_i/d\Omega$ at 
40$^{\circ}$ was found constant to within 20\% for the 
different excited states.

The mean excitation energy for $E_p$ = 45 MeV is then simply defined as:

\begin{equation} {\mathcal E}_x = \frac{\sum_{i}~ P_i~E_i}{\sum_{i}~P_i}
\end{equation}

To obtain the mean excitation energy at other proton energies, we made
extensive OM calculations with the code Ecis94 \cite{Ray}, using the optical
potential parameters of the compilation of Perey and Perey \cite{Per}. For each
proton energy we calculated the inelastic scattering cross section for the
eight representative states of $^{16}$O. All calculations were restricted to
direct one-step excitation of the respective state in $^{16}$O, which should
be the main excitation mechanism for the states of interest at these proton
energies.

Calculations were done for the following proton energies: $E_p$ = 
22.5, 25, 30, 40, 45, 66 and 100 MeV. The mean excitation energy as a
function of proton energy was calculated in the following way,

\begin{eqnarray}  {\mathcal E}_x (E_p)~ = ~ \frac{\sum_{i}~ P_i ~ E_i ~
\sigma_i(E_p)/\sigma_i(45 ~MeV)}{\sum_i ~ P_i ~
\sigma_i(E_p)/\sigma_i(45 ~MeV)} 
\end{eqnarray}

The differential cross sections have been obtained in the same way by
averaging over the eight representative states.  In order to facilitate line
shape calculations for a wide range of proton energies, we parametrized the
differential cross section by a simple function:

\begin{equation} \frac{d\sigma}{d\Omega_p}~~ = ~~ N ~ e^{-p ~\Theta} 
\end{equation}

where $\Theta$ is the proton diffusion angle in the center-of-mass system.
It was checked that this simple function produced essentially the same line
shapes as the differential cross sections from the OM calculations.  The
values of $\mathcal E$$_x$ and {\it p} are presented in Table \ref{tab1}.

\section{The inelastic scattering reaction $^{12}C(\lowercase{p,p}\gamma)$ }

The cross section excitation function $\sigma(E_p)$ was taken from the
measurements of Dyer et al. \cite{Dyer} from threshold to 23 MeV, above that
from the data of Lang et al. \cite{LW} and Lesko et al. \cite{Les} up to 85
MeV, then from the curve of Ramaty, Kozlovsky and Lingenfelter \cite{RKL} up
to 100 MeV. Cross section values are reported in Table \ref{tab2}.

Production of 4.438 MeV $\gamma$-ray emission by inelastic scattering off
$^{12}$C originates mainly from direct excitation of the 2$^+$, 4.439 MeV
level. All higher-lying levels of $^{12}$C are particle-unbound and have
small to very small $\gamma$-branching ratios to the 4.439 MeV state
\cite{Fir}, making $\gamma$-ray cascade contributions negligible. The
$\gamma$-ray line shape and the $\gamma$-ray angular distribution are in
this case completely determined by the double differential inelastic cross
section d$^2\sigma$/d$\Omega_p$d$\Omega_{\gamma}$. It can be expressed as a
product of the differential cross section d$\sigma$/d$\Omega_p$ and the
angular correlation function W (see for example Satchler \cite{Sat}):

\begin{equation} \frac{d^2 \sigma}{d \Omega_p d \Omega_{\gamma}} ~~ = ~~
\frac{d \sigma}{d \Omega_p} ~ \frac{1}{4 \pi} ~ 
W(\vec{k},\vec{k}',\vec{k}_{\gamma}) 
\label{eqw}
\end{equation}

where $\vec{k}$ and $\vec{k}'$ are the wave vectors of the incoming and
outgoing proton, respectively, and $\vec{k}_{\gamma}$ is the vector of the
emitted $\gamma$-ray.

Although a wealth of differential cross sections are available for this
reaction, proton-$\gamma$ correlation measurements exist only for few proton
energies. The angular correlation function must therefore be extracted
either from experimental data, for example by fitting of $\gamma$-ray
angular distributions and line shapes or be obtained from OM calculations.
The excitation function for the 4.438 MeV $\gamma$-production cross section
exhibits pronounced fluctuations from threshold up to $\approx$ 15 MeV (see
Ref. \cite{KB} or Ref. \cite{Dyer}). This indicates strong contributions of
compound nucleus resonances $^{12}$C+p $\rightarrow$ $^{13}$N$^{\star}$
$\rightarrow$ p+$^{12}$C$^{\star}_{4.439}$ to the inelastic scattering.

Because OM calculations are only suited for the direct reaction mechanism,
their use in this energy range was discarded. We used the measured line
shapes and $\gamma$-ray angular distributions of the Orsay experiment
\cite{KB} and published data for
d$\sigma$/d$\Omega_p$ for $E_p$ = 8.6 - 20 MeV to adjust the parameters of
the correlation function. Some available $\gamma$-ray angular distributions 
of the Washington experiment \cite{Dyer} were also used for comparison. 
The parameter adjustment was achieved by reproducing as closely as
possible the experimental data in extensive calculations with the
Monte-Carlo type program simulating the reaction and the detection setup of
the Orsay experiment.

At higher proton energies, Meigooni et al. \cite{Mei} provide a
phenomenological optical potential for neutron and proton scattering off
$^{12}$C, which reproduces fairly well elastic and inelastic scattering
angular distributions. The OM calculations were used to obtain the necessary
input parameters for the calculation of the double differential cross
section d$^2 \sigma$/d$\Omega_p$d$\Omega_{\gamma}$.

\subsection{$E_p$ = 8.6 - 20 MeV}

We adopted the formalism used in Refs. \cite{RKL} and \cite{KRL} for the
proton-$\gamma$ angular correlation and the calculation of the $\gamma$-ray
line shape. In this formalism, the angular correlation function W is
expressed in the rest frame of the recoiling excited
$^{12}$C$^{\star}_{4.439}$ nucleus, whose recoil angle and velocity is
kinematically fixed by the incoming and outgoing proton wave vectors
$\vec{k}$ and $\vec{k}'$. The correlation function W then reduces to the
$\gamma$-ray angular distribution $W(\Theta,\Phi)$ in the carbon rest frame.
It is proportional to the radiation pattern $dP/d\Omega$, which depends on
the magnetic substate population amplitudes $a_m(\vec{k},\vec{k}')$ of the
4.439 MeV, 2$^+$ state of $^{12}$C (see e.g. Jackson \cite{Jac}):

\begin{equation} \frac{dP}{d\Omega} ~~ = ~~ \frac{c}{8 \pi k^2} ~
\mid  \sum_{m=-2}^{2}  i~a_m(\vec{k},\vec{k}') ~ \vec{X}_{2m}(\Theta, \Phi) 
\mid^2 
\end{equation}

where the $\vec{X}_{2m}$ are the vector spherical harmonics for quadrupole
radiation, and $\Theta$, $\Phi$ the polar and azimuthal angles of the
$\gamma$-ray emission in the rest frame of the excited carbon. The aim was
to find a set of population amplitudes $a_m(\vec{k},\vec{k}')$ for each
proton energy of the Orsay experiment, which gives simultaneously a good
reproduction of the experimental line shapes and the laboratory $\gamma$-ray
angular distribution.

Kozlovsky, Ramaty and Lingenfelter \cite{KRL} obtained reasonable fits of
the six experimental line shapes at $E_p$ = 23 MeV from ref. \cite{KAG} by
adjusting only two independent real amplitudes, which furthermore were
independent of the proton diffusion angle. We started our search with the
same simple basis and adjusted the two independent relative amplitudes
$a_2$/$a_0$ and $a_1$/$a_0$ (with $a_m$ real, $a_{-m}$ = $a_m$ and $\sum
a_m^2$ = 1), independent of the carbon recoil angle. However, these two
parameters were not sufficient and it proved to be necessary for a
simultaneous reproduction of the line shapes and the $\gamma$-ray angular
distribution to introduce a third independent parameter. We found that a
phase shift $\Delta\Phi$ for the azimuthal angle improved considerably the
fits and gave a satisfactory reproduction of the Orsay data at most proton
energies.

Practically, for the search of the population amplitudes, we used as in the
case of the spallation reaction a Monte-Carlo type program simulating the
nuclear reaction and the detection system of the Orsay experiment. For each
proton center-of-mass diffusion angle, the $^{12}$C$^{\star}_{4.439}$ recoil
direction and energy were calculated in the laboratory system. For the
$\gamma$-ray emission, we chose a coordinate system with origin in the
recoiling carbon nucleus, the z-axis perpendicular to the scattering plane
and the x-axis in the direction of the carbon velocity vector. $\Theta$
defines the angle between the $\gamma$-ray vector and the z-axis, while
$\Phi$ is defined as the angle between the x-axis and the projection of the
$\gamma$-ray vector on the scattering plane, counted counterclockwise from
the x-axis.  Subsequently, the $\gamma$-ray vector was calculated in the
laboratory system and the energy of the $\gamma$-ray was stored when it was
in the solid angle of one of the detectors. Again, slowing down of the
recoiling excited $^{12}$C in the target and decay of the 4.439 MeV state
with half-life $\tau_{1/2}$ = 42 fs were taken into account.

Because of the presence of multiple compound nucleus resonances in this
energy range, calculations were done in narrow proton energy steps between
8.6 and 20 MeV. For each energy, population amplitudes were adjusted to
reproduce the $\gamma$-ray line shapes of the six detectors and special care
was taken to obtain  a simultaneously good reproduction of the
laboratory $\gamma$-ray angular distribution. For d$\sigma$/d$\Omega_p$,
experimental cross sections of Peelle \cite{Pel} for $E_p$ = 14 - 19.4 MeV
and of Barnard, Swint and Clegg \cite{BSC} for $E_p$ = 8.6 - 11.6 MeV were
used. For $E_p$ = 11.6 - 14 MeV we interpolated between the differential
cross sections of Ref. \cite{BSC} at 11.6 MeV and Ref. \cite{Pel} at 14 MeV.

The results of the adjustments are presented in Table 
\ref{tab2} and an example of
calculated line shapes and laboratory $\gamma$-ray angular distribution is
shown in Fig.\ \ref{fig2} together with data of the Orsay \cite{KB} and
Washington \cite{Dyer} experiment.

\subsection{$E_p$ = 20 - 100 MeV}

Experimental data in this energy domain are relatively scarce, and no
systematic measurements of differential cross sections d$\sigma$/d$\Omega_p$
and d$\sigma$/d$\Omega_{\gamma}$ or line shapes are available. However,
Meigooni et al. \cite{Mei} performed a systematic study of nucleon elastic and
inelastic scattering off $^{12}$C and found an energy dependent optical
potential parameter set, which provides good fits to measured differential
cross sections in a wide angular range and for energies above the pronounced
compound nucleus contributions ($\approx$ 15 MeV) up to approximately 100
MeV. We therefore based our calculations entirely on OM calculations with
their parameter set.

These calculations were done with the program Ecis94 \cite{Ray} to obtain
the scattering amplitudes $T_{M_A',M_a',M_A,M_a}$($\vec{k}, \vec{k}'$)
necessary for the construction of the correlation function W
(Eq.\ \ref{eqw}). $M_A$ ($M_A'$)
and $M_a$ ($M_a'$) are the z-components of the $^{12}$C and the proton spin 
in the incoming (outgoing) channel, respectively. 
  Choosing the z-axis along the incoming proton direction $\vec{k}$,
the amplitudes, correlation function and consequently the proton and
$\gamma$-emission angle are directly given in a space-fixed system with
respect to the beam direction of the experiment. The correlation function
$W$ for emission of the $\gamma$-ray in the direction $\vec{k}_{\gamma}$ is
given by:

\begin{equation} W(\vec{k},\vec{k}',\vec{k}_{\gamma}) ~~ = ~~ \sum_{kq} ~ 
t_{kq}(\vec{k},\vec{k}')  ~
R_{kq}^{\star} \end{equation}

where $t_{kq}$ denotes the polarization tensor. It is constructed from the
scattering amplitudes $T$ and contains all information on the state of
polarization of the excited carbon after scattering (see Satchler \cite{Sat}, 
section 10.3.3). $R_{kq}$ are the radiation tensors:

\begin{equation} R_{kq} ~~ = ~~R_k(\gamma) ~ \sqrt{\frac{4\pi}{2k+1}} ~
Y_k^q(\Theta_{\gamma},\Phi_{\gamma}) \end{equation}

where $R_k(\gamma)$ are the gamma radiation parameters and $Y_k^q$ the
 spherical harmonics.

Introducing this correlation function in a program similar to the
Monte-Carlo type program described above, we calculated theoretical line
shapes and $\gamma$-ray angular distributions in the laboratory, using the
differential cross sections d$\sigma$/d$\Omega_p$ obtained from OM 
calculations.
Comparison with the Orsay data for $E_p$ = 16-20 MeV gave fairly good
reproductions of the line shapes and $\gamma$-ray angular distributions. An
example is shown in Fig.\ \ref{fig3} for $E_p$ = 19.75 MeV. 
Given the tendency that
the differential cross sections are better reproduced with increasing proton
energy (see Ref. \cite{Mei}), the 
calculated line shapes and $\gamma$-ray angular
distributions should be fairly realistic above 20 MeV.

Those calculations are, however, quite complex, involving for each proton
energy the input of five complex scattering amplitudes depending furthermore
on the proton center-of-mass scattering angle. In order to facilitate line
shape calculations for proton spectra covering a wide energy range as in
solar flares or cosmic rays, the line shapes and $\gamma$-ray angular
distributions from the OM calculations for $E_p$ $>$ 20 MeV, were adjusted
with the population amplitude method described in subsection III.A.  The
population amplitudes and phase shifts from these adjustments are reported
in Table \ref{tab2}.

\section{alpha-particle induced reactions}

As in the case of proton induced reactions, the main source of 4.438 MeV
$\gamma$-rays by accelerated $\alpha$-particles is inelastic scattering off
$^{12}$C and spallation of $^{16}$O. The cross section excitation functions
being quite similar for both projectiles, reactions with
$\alpha$-particles should generally not contribute significantly to the
$\gamma$-ray line due to their low abundance, e.g. an observed $\alpha$/p
ratio of 0.035 and 0.0076 in the energetic particle spectrum of two solar
flares \cite{MDR}.

However, $\alpha$/p ratios in excess of 0.1 for the accelerated particle
spectrum were shown to be necessary for an acceptable fit of the
$\gamma$-ray spectrum of the 1981 April 27 flare \cite{MRK}.
$\alpha$-particle reactions are furthermore favored in the case of a very
soft particle spectrum, because the thresholds for inelastic scattering off
$^{12}$C and spallation of $^{16}$O are lower for $\alpha$-particles
($\approx$ 2 and 5 MeV per nucleon, respectively) than for protons
($\approx$ 6 and 14 MeV, respectively). In such cases, an important fraction
of the $\gamma$-ray production may be due to reactions with
$\alpha$-particles. Therefore, a treatment of $\alpha$-particle reactions
similar to the above described studies on proton reactions seems worthwhile.

Unfortunately, experimental line shapes are practically absent for
$\alpha$-particle inelastic scattering off $^{12}$C and do not exist for
$\alpha$-particle induced spallation of $^{16}$O. Concerning spallation of
$^{16}$O, this may not be very problematic because of the excitation of many
intermediate levels in $^{16}$O resulting in essentially isotropic
$\gamma$-ray angular distributions as for $^{16}$O(p,p$\alpha \gamma$).
Optical model calculations should then provide as for proton induced
spallation a reasonable estimation of the mean excitation energy in $^{16}$O
and approximate differential inelastic cross sections. On the contrary,
inelastic $\alpha$-particle scattering off $^{12}$C is certainly dominated
at energies below $E_{\alpha}$ $\approx$ 30 MeV by compound nucleus
resonances due to the pronounced $\alpha$-particle structure of $^{12}$C and
$^{16}$O, making OM calculations very hazardous in this energy range.

\subsection{The spallation reaction $^{16}$O($\alpha$,x$\gamma$)$^{12}$C}

We made the same assumptions as for the case of proton induced spallation of
$^{16}$O. Only the sequential process of excitation of intermediate levels
in $^{16}$O with subsequent decay in $\alpha$ and $^{12}$C$^{\star}_{4.439}$
was considered. The angular distributions of the excited carbon in the
system of the excited oxygen and of the $\gamma$-ray in the system of the
excited carbon were both considered isotropic.

Because of the absence of experimental line shapes, the estimation of the
mean excitation energy in $^{16}$O was entirely based on OM calculations and
the $\alpha$-particle energy spectrum of Lang et al. \cite{LW}, obtained at
$E_{\alpha}$ = 60 MeV and at a scattering angle of 12.5$^{\circ}$. Contrary
to proton scattering, the ratio of the total inelastic scattering cross
section to the differential cross section at 12.5$^{\circ}$ shows a
systematic dependance on the excitation energy in the considered range $E_x$
= 12.53 - 28 MeV. Therefore, the relative contributions
$d\sigma_i/d\Omega(12.5^{\circ})$ extracted from the spectrum of
Ref. \cite{LW}, were multiplied by this ratio to obtain the relative inelastic
scattering cross sections $\sigma_i$ at $E_{\alpha}$ = 60 MeV. Finally,
for the partial $\gamma$-production probability P$_i$ at 60 MeV, the
inelastic scattering cross sections were multiplied with the branching
ratios $\Gamma^i_{\alpha_1}$/$\Gamma^i_{tot}$, proposed by Ref.
\cite{LW}.

Extensive OM calculations were done for $\alpha$-particle energies
$E_{\alpha}$ = 18 - 400 MeV and the same eight representative excited
$^{16}$O-states as in subsection II.B to estimate their relative contribution
P$_i$ at other projectile energies. As in the proton case, we restricted the
calculations to direct one-step $l=2$ excitations. The energy-dependent
phenomenological optical potential of Michel et al. \cite{Mich}, which gives
good to very good fits of elastic differential cross sections from about 20
MeV to 146 MeV, were used in the entire energy range.  From the same
calculations, averaged differential cross sections were constructed and
their angular distributions fitted by a parameterization as in subsection
II.B, but with the addition of a term cos$^4$($\Theta$) to take account of
the backward enhancement of the differential cross sections:

\begin{equation}
\frac{d \sigma}{d \Omega} ~ = ~ C_1 e^{-p \Theta} + C_2 cos^4(\Theta)
\end{equation}

For this reaction, the $\gamma$-ray production cross section has only been
measured at five $\alpha$-particle energies by
Dyer et al. \cite{Dyer2} from 22 to 26 MeV and by Zobel et al.
\cite{Zob} at $E_{\alpha}$ = 52 MeV. An estimated cross section curve is
given in Ref. \cite{RKL}, however based on the then only available cross 
section data of Ref. \cite{Zob} and limited to $\alpha$-particle energies 
below 20 MeV
per nucleon. We therefore made a new estimation extending the energy range
from threshold to 100 MeV per nucleon. Our best guess, based on the energy
dependence of the ($\alpha$,4pxn) spallation data of Lange et al.
\cite{Lange} and  the OM calculations is a linear interpolation
between the cross sections at 26 MeV and 52 MeV. Below 22 MeV and above 52
MeV, we take the energy dependence of the OM calculations.

The results of the calculations for the mean excitation energy, the
inelastic scattering angular distributions and the $\gamma$-ray production 
cross sections are presented in Table \ref{tab3}.  

\subsection{The inelastic scattering reaction 
$^{12}$C($\alpha$,$\alpha\gamma$)}

As already mentionned, practically no line shapes are available for this
reaction and OM calculations at low $\alpha$-particle energies
are not believed to be very realistic. To get anyway an idea of the magnetic
substate population amplitudes, we employed two different methods at energies
below 30 MeV. From $E_{\alpha}$ = 8 - 17 MeV, we used the systematic
measurements of differential cross sections d$\sigma$/d$\Omega_{\alpha}$ and
laboratory $\gamma$-ray angular distributions d$\sigma$/d$\Omega_{\gamma}$
of Mitchell, Carter and Davis \cite{Mit} and Ophel et al.\cite{Oph}. We then
used the Monte-Carlo type program described in subsection III.A to find a set
of population amplitudes $a_m$ (with $a_m$=$a_{-m}$, and here throughout 
$a_{\pm 1}$=0 for
0$^+$ $\rightarrow$ 2$^+$ excitations induced by
inelastic $\alpha$-particle scattering on spin-0 targets like $^{12}$C)
and phase shifts $\Delta \Phi$ which reproduces the measured 
$\gamma$-ray angular distributions.

Morgan and Hobbie \cite{Mor} measured in small energy steps between
$E_{\alpha}$ = 19-30 MeV inelastic differential cross sections. However, no
$\gamma$-ray angular distributions or line shape measurements are available
in that energy range. Therefore, we used OM calculations despite the fact
that compound nucleus resonances are probably important below $E_{\alpha}$ =
30 MeV. In order to reduce the effect of individual compound nucleus
resonances, we averaged several differential cross sections of Ref.
\cite{Mor} centered around the $\alpha$-particle energies of 21 and 25 MeV.
Between 30 MeV and 200 MeV, we used the inelastic scattering data of Burdzik
and Heymann at 32.5 MeV \cite{Bur}, of Baron, Leonard and Stewart \cite{BLS}
at 41 MeV, of D'Agostino et al. at 90 MeV \cite{dago}, of Smith et al. at
139 MeV \cite{Smi} and of Tatischeff and Brissaud at 166 MeV \cite{TB}. With
these differential cross sections OM fits were then done, again with the
optical potential of Michel et al \cite{Mich}. Results of the OM fits in the
energy range $E_{\alpha}$ = 17 - 166 MeV are shown in Fig.\ \ref{fig4}.
Above 200 MeV, no inelastic differential cross sections are published. We
continued to use the optical potential shape of Michel et al. for the
inelastic scattering calculations, extrapolating up to 400 MeV the optical
potential found at 166 MeV by following the energy dependence of the
potential parameters given by Michel et al. for $\alpha$-particle scattering
off $^{16}$O.

Again, as for proton inelastic scattering off $^{12}$C, the
line shapes and $\gamma$-ray angular distributions
predicted by the OM calculations were used to deduce population
amplitudes and phase shifts for a consistent parameterization of the
$\gamma$-ray emission in proton and $\alpha$-particle inelastic scattering 
off $^{12}$C in the energy range of interest for solar flares and low energy
cosmic rays. The cross section
excitation function is taken from Dyer et al. \cite{Dyer2} for $E_{\alpha}$ =
8-26 MeV, from the data of Refs. \cite{Bur,BLS,dago,Smi,TB} 
at 32.5, 41, 90, 139 and 166 MeV, respectively, and from our OM
calculations at 250 and 400 MeV. Results for the population amplitudes,  
phase shifts and total cross sections are presented in Table \ref{tab4}. 

\section{Predicted line shapes in solar flares and discussion}

With the above deduced reaction parameters, the line shape of the 4.438 MeV
$\gamma$-ray emitted from the first excited state of $^{12}$C for proton and
$\alpha$-particle interactions with $^{12}$C and $^{16}$O can be calculated
for energies from threshold up to 100 MeV per nucleon. To illustrate the
line shapes that can be expected from solar flares, we use a simple model
of $\gamma$-ray production by energetic particle interactions in a thick
target. For the accelerated particle energy distribution $N(E)$, we choose a
spectrum as it results from diffuse shock acceleration \cite{MDR}:

\begin{equation} N(E)  ~ \propto ~ \frac{1}{v} ~ p^{-s} ~ exp(-E/E_0)
\label{eqne}
\end{equation}

where $v$ and $p$ are the particle velocity and the momentum per nucleon,
respectively; $s$ and $E_0$ are the spectral index and cutoff energy,
specific to a given acceleration site. The energies labelled $E$, $E_0$,
$E_{max}$, $E'$ in Eqs.\ (\ref{eqne},\ref{eqng})
and in the following are expressed in energy
per nucleon. The total number of 4.438 MeV $\gamma$-rays $N_{\gamma}$
produced by interactions of projectiles of type $i$ (p or $\alpha$) with 
target nuclei of type $j$ is then given by:

\begin{equation} N_{\gamma}(E) ~~ = ~~ [A_j]~n~  \times \int_0^{E_{max}}
\frac{\sigma_{ij}(E)dE}{dE/dx_i(E)}
\times \int_{E}^{E_{max}}~ N_i(E') dE'~ 
\label{eqng}
\end{equation}

where $[A_j]$ is the abundance of isotope $j$ with respect to hydrogen and $n$
the  density of hydrogen atoms
at the interaction site; $\sigma_{ij}(E)$ is the $\gamma$-ray production
cross section and $dE/dx_i(E)$ is the stopping power of particle $i$.
We made the simplifying assumption that the energy loss is only due to the
electronic stopping power in a neutral gas, consisting of hydrogen and
helium with $[He]/[H]$ = 0.1. The Bethe formula was used for the energy loss
calculation. 

We adopted two sets of $s$ and $E_0$: $s$ = 3.3;
$E_0$ =
30 MeV and $s$ = 2.4; $E_0$ $\geq$ 300 MeV  corresponding to
the fit of interplanetary proton observations of two distinct solar flares
\cite{MDR}. We set $[O]$ = 0.068\%, $[C]$ = 0.042\% as in Ref. \cite{MDR},
$E_{max}$ = 2 $\times$ $E_0$ and an  $\alpha$/p-ratio of 0.1. 

In both cases, proton interactions dominate; they account for 82\% and 84\%
of the $\gamma$-ray production with the softer and harder spectrum,
respectively.
With the softer spectrum, the $\gamma$-ray production is dominated
by proton inelastic scattering off $^{12}$C (57\%), of which in
particular 85\% of the $\gamma$-rays are produced by interactions at proton
energies below 20 MeV.
In order to get an overall impression of 
the quality of reproduction of 
the $\gamma$-ray lines from the important low energy proton inelastic 
scattering reactions, energy integrated data from the Orsay experiment 
were produced and compared with  the calculations.
Line shapes and $\gamma$-ray angular distributions for energy integrated
experimental and calculated data between 8 MeV and 20 MeV for proton
inelastic scattering off $^{12}$C, are represented in Fig.\ \ref{fig5}.    

For the relatively hard spectrum with $s$ = 2.4 and setting $E_0$
= 300 MeV, proton induced spallation of
$^{16}$O is the strongest channel, accounting for 46\% of the 4.438 MeV
$\gamma$-rays. Here, proton interactions at energies below 20 MeV are 
responsible for 24\% of the total $\gamma$-ray production, $\alpha$ and
proton interactions at energies below 100 MeV account for 83\%
of the produced $\gamma$-rays. These values, however, have to be regarded as
lower limits only, because secondary protons of low and intermediate energy 
produced by
high energy proton and $\alpha$-particles and escape of accelerated
particles into interplanetary space are not taken into account.

For the prediction of $\gamma$-ray line profiles from solar flares
 we calculate the interaction probability as in Eq.\ (\ref{eqng}).
 No absorption or Compton scattering
of the produced 4.438 MeV $\gamma$-ray in the solar atmosphere is taken
into account. Three different energetic particle angular distributions have
been investigated, similar to the distributions investigated by Murphy,
Kozlovsky and Ramaty \cite{MKR}: an isotropic angular distribution, a
fan-beam flare and a downward flare. The narrow component of the 4.438 MeV
$\gamma$-ray is shown in Figs.\ \ref{fig6}, \ref{fig7} and \ref{fig8} 
for the different energetic particle
angular distributions and energy spectra, all with an $\alpha$/p-ratio 
of 0.1. All curves are arbitrarily normalized to a total kinetic
energy in proton and $\alpha$-particles with $E>$1 MeV of 10$^8$ erg.

 The narrow component is clearly dominated by the p+$^{12}$C reaction
for the soft particle spectrum. Comparing curves 1 and 2 of
Fig. \ref{fig6},
which shows the predicted line shapes for isotropic incident particle
spectra, it seems possible to estimate the spectral hardness by the
width of the line  and the characteristic drop around 4.438 MeV only
visible for the hard spectrum. This drop, also seen in curve 5 of Fig.\ 
\ref{fig6} is not predicted by other parameterizations, as for example in Ref. 
\cite{MKR}.

Figure \ref{fig7}
shows the line profiles resulting from a downward directed energetic particle
distribution at the solar limb. The parameterization of Ref. \cite{MKR} for
the population amplitudes reproduces qualitatively the characteristic 
line shape with two intensity maxima above and below the minimum at 4.438 MeV,
but it
underestimates slightly the maximum-to-minimum ratio and overestimates
the intensity (by $\approx$10\%) emitted at 90$^{\circ}$. Line-shapes
produced by fan-beam distributions and an isotropic distribution as in 
Fig.\ \ref{fig8} 
will be difficult to distinguish, showing only small differences in the
overall line profile.

In summary, with the extracted nuclear reaction parameters in this work,
line shape calculations for proton and $\alpha$-particle induced reactions on
$^{12}$C and $^{16}$O can be done for all possible energetic particle
angular distributions and for particle energies from threshold to about
100 MeV per nucleon, which is the important energy domain for solar flares
and low energy cosmic rays.
These parameters can also be straightforwardly used for the inverse
reactions, energetic $^{12}$C and $^{16}$O bombarding proton and
$\alpha$-particles. 

\begin{figure}
\epsfig{file=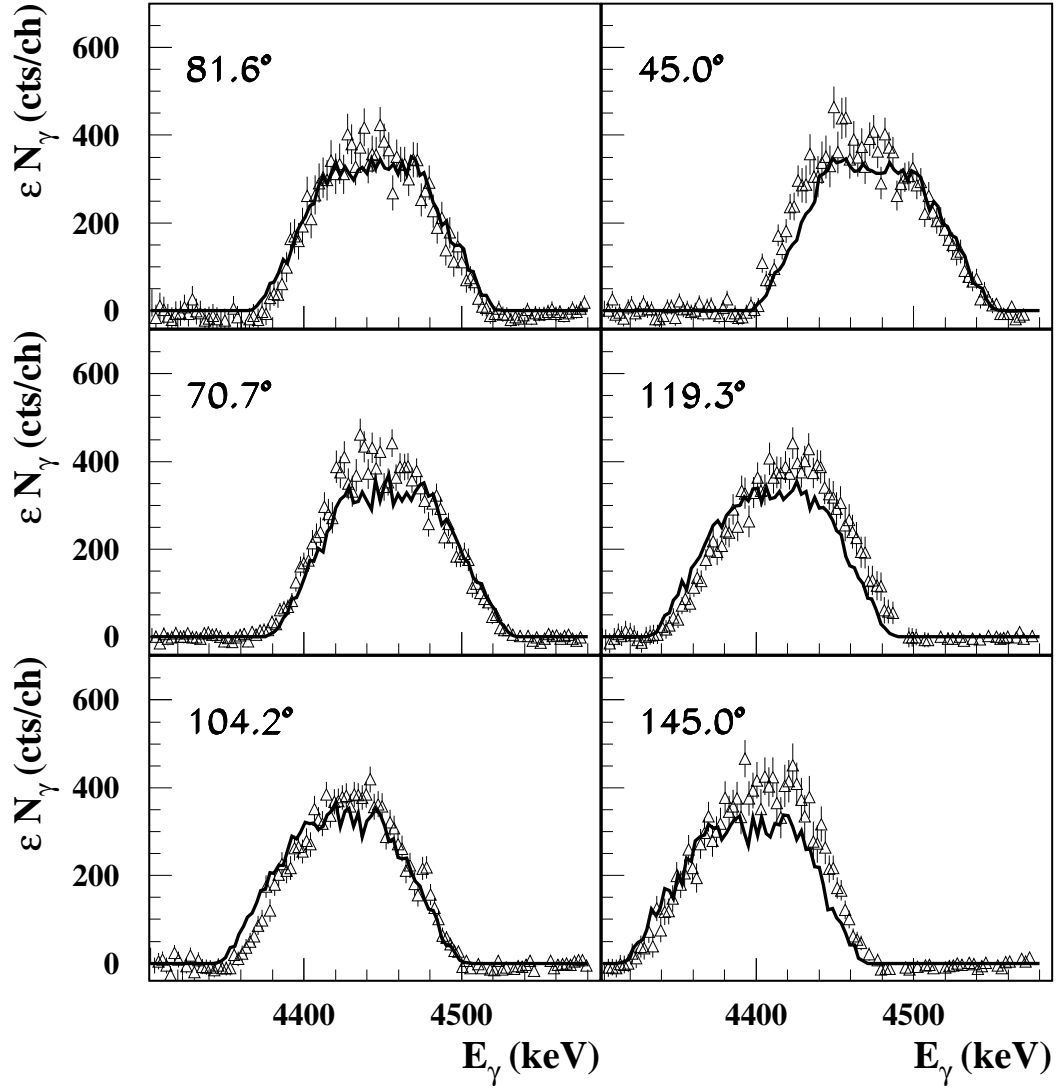}
\caption{
Experimental spectra from the Orsay experiment [8] of the 4.438 MeV
$\gamma$-ray line (open triangles) from the proton induced spallation of
$^{16}$O at $E_p$ = 20 MeV and results of a Monte-Carlo simulation of the
experiment (full line) with a constant differential cross section in the
center-of-mass proton diffusion angle and a mean excitation energy of $E_x$
= 14.75 MeV. Note that the experimental spectra at the different detector
angles have been corrected for the individual relative detector
efficiencies. The fact that the simulation with isotropic angular
distributions predicts correct relative line intensities (with a common
absolute normalization factor) indicates therefore an isotropic angular
distribution of the $\gamma$-rays in the laboratory.}
\label{fig1}
\end{figure}

\begin{figure}
\epsfig{file=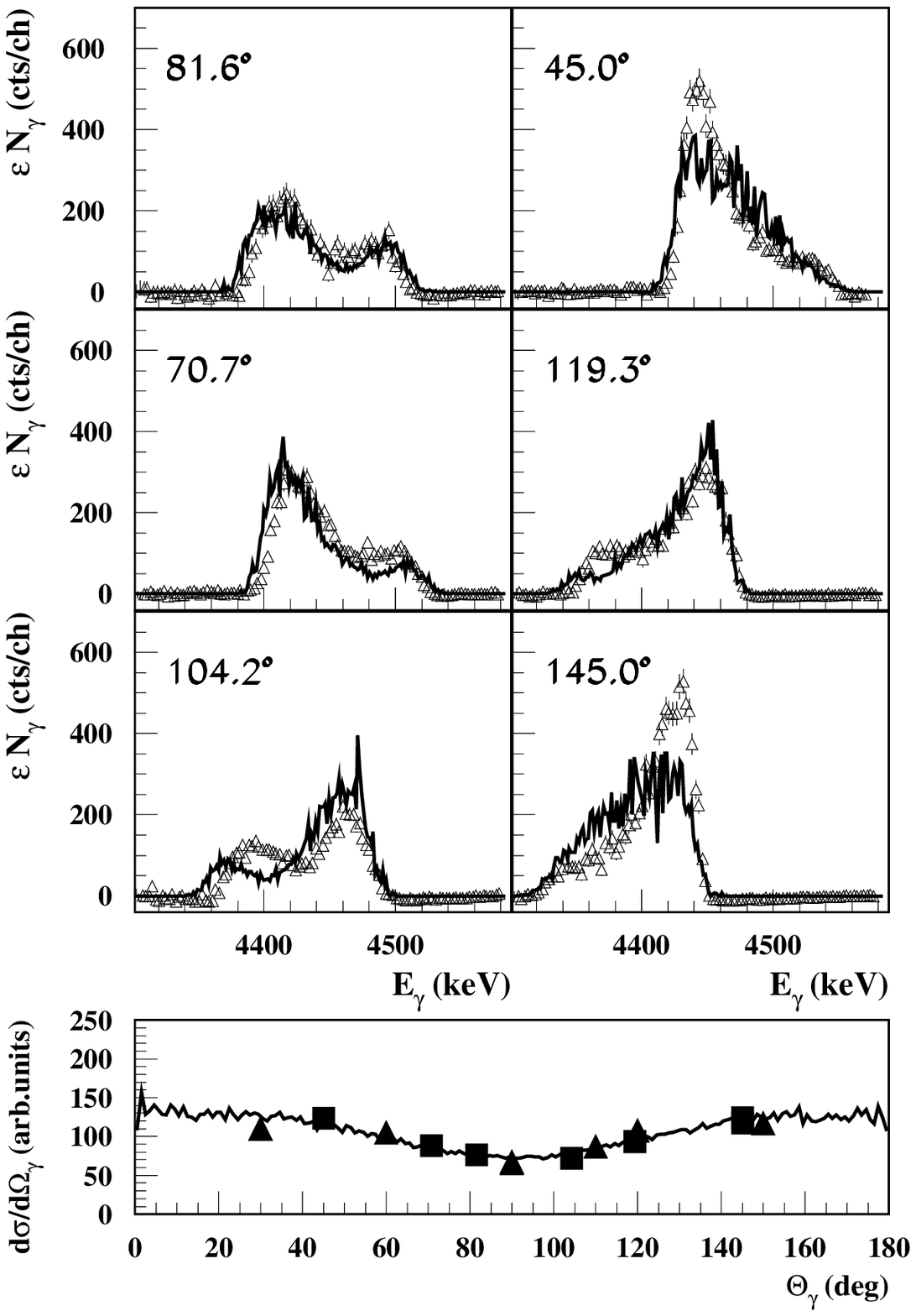}
\caption{ (Upper part): Line shapes for the 4.438 MeV $\gamma$-ray at
six different detection angles from proton inelastic scattering off $^{12}$C
at $E_p$ = 19.75 MeV. Triangles: count spectra of the Orsay experiment [8],
full line: calculated line shapes with the population amplitude method.
\newline
(Lower part): Laboratory $\gamma$-ray angular distribution from that 
reaction. Full squares: data from the Orsay experiment, full
triangles: experimental data of the Washington experiment [11] 
at $E_p$ = 20 MeV, 
normalized to the Orsay data, full line: calculated distribution with the
population amplitude method.}  
\label{fig2}
\end{figure}

\begin{figure}
\epsfig{file=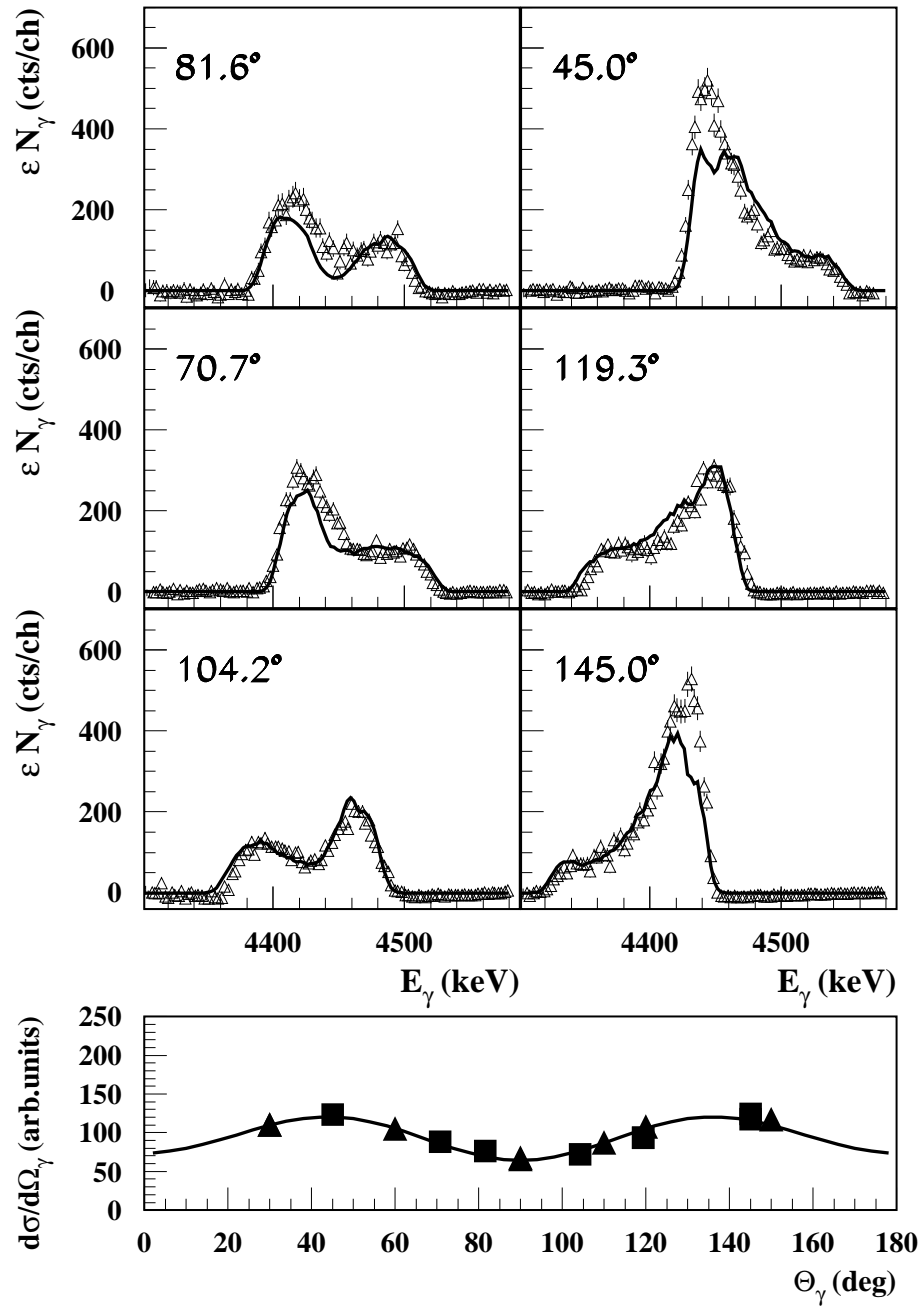}
\caption{
Same as Fig.\ \ref{fig2},  except
full lines: line shapes and $\gamma$-ray angular distribution 
resulting from OM calculations.}  
\label{fig3}
\end{figure}

\begin{figure}
\epsfig{file=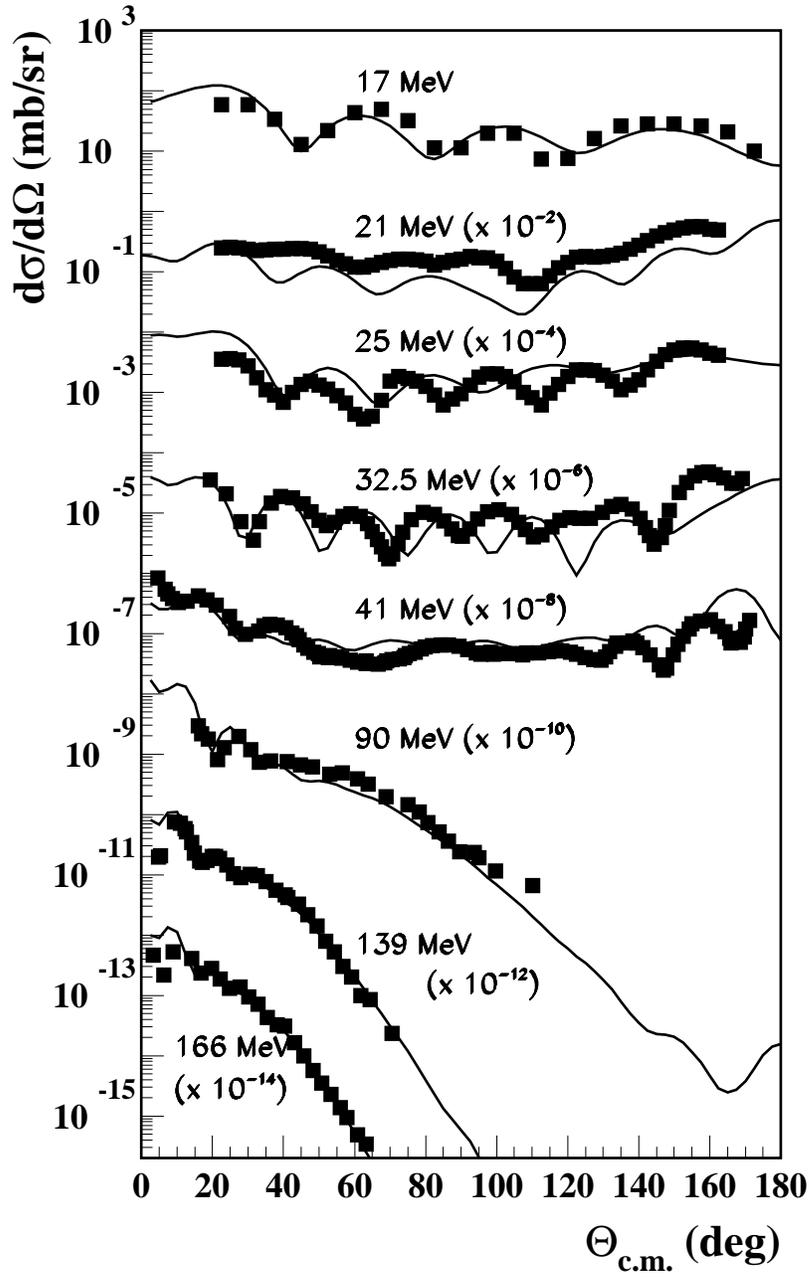}
\caption{
Experimental differential inelastic cross sections 
$^{12}$C($\alpha$,$\alpha'$)$^{12}$C (filled squares, Refs.
[31-37]) and results of OM 
fits (full lines).}
\label{fig4}
\end{figure}

\begin{figure}
\epsfig{file=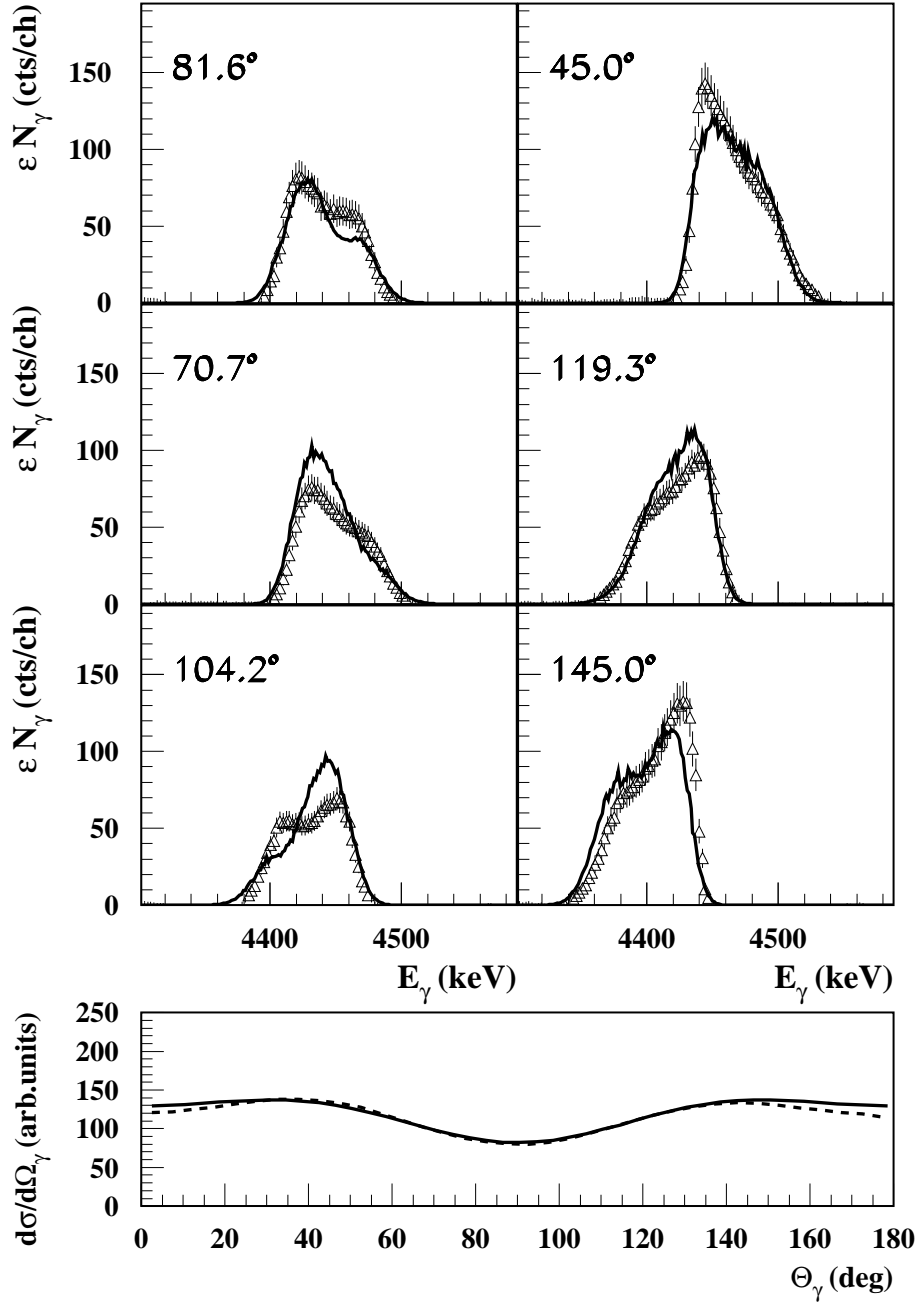}
\caption{
 (Upper part): Line shapes for the 4.438 MeV $\gamma$-ray
from proton inelastic
scattering off $^{12}$C summed for $E_p$ = 8.6-20 MeV.
The summation has been done with a weight factor for each
proton energy following the thick-target interaction probability of a
incident proton spectrum as given in Eq.\ (\ref{eqne}). Parameters of the
spectrum were $s$=3.3; $E_0$=30.
The summed experimental spectra from the Orsay experiment [8] are
presented by open triangles while the summed spectra constructed with the
population amplitudes method are shown by the full line. 
\newline
(Lower part): Summed laboratory $\gamma$-ray angular distributions. 
The summed
experimental distributions are constructed from Legendre-polynom fits of 
the Orsay data  and are shown by the full line, while the dashed 
line shows the angular
distribution resulting from the population amplitudes method.}
\label{fig5}
\end{figure}

\begin{figure}
\epsfig{file=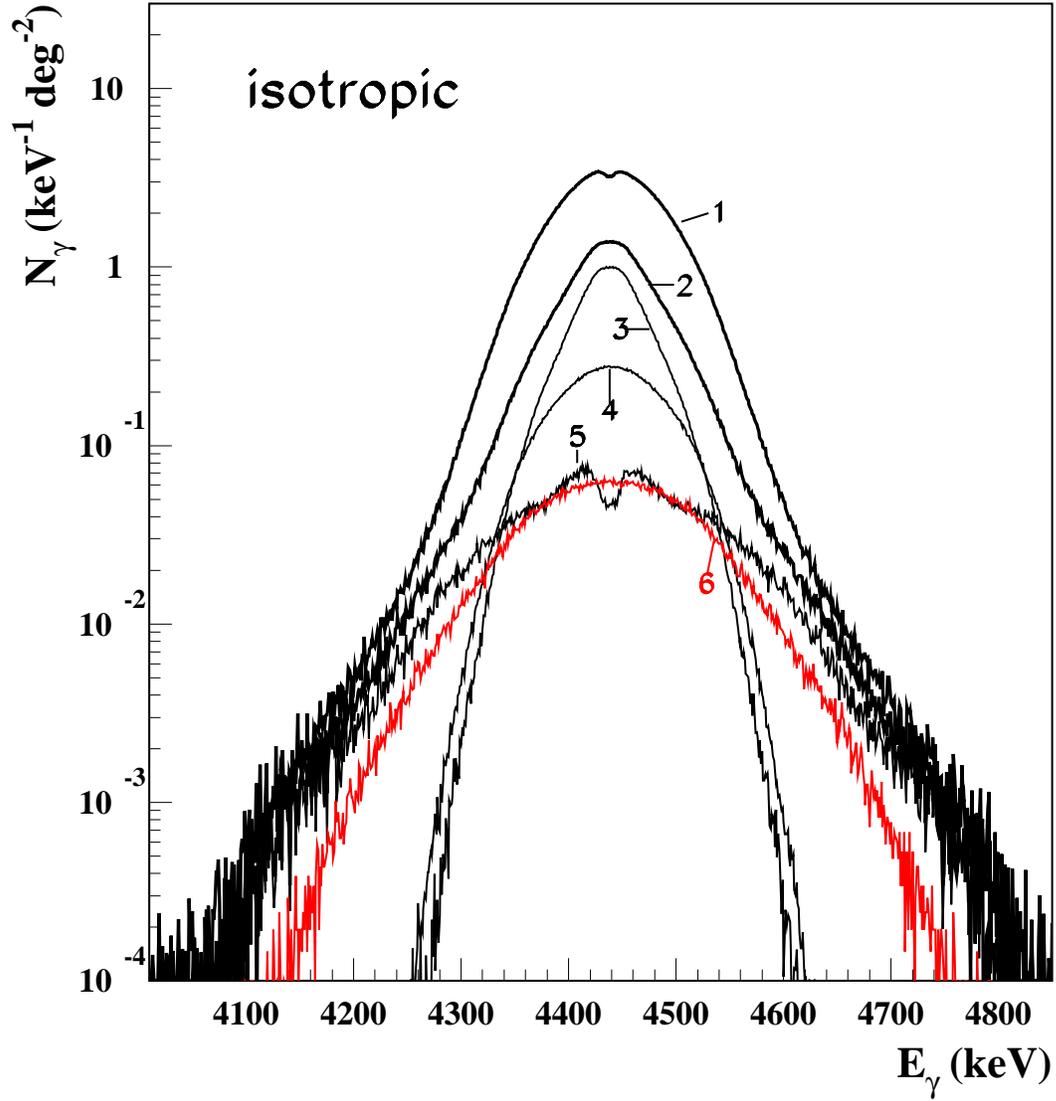}
\caption{
Emitted line profiles of the 4.438 MeV $\gamma$-ray for flares with
isotropic energetic particle distribution.
Curves labelled 1 and 2 are the profiles resulting
from energetic proton and $\alpha$-particle interactions with $^{12}$C and
$^{16}$O with spectral parameters $s$=2.4; $E_0$=300 MeV and $s$=3.3; 
$E_0$=30 MeV, respectively.  Curves
labelled 3-6 show the different contributions for the soft particle
spectrum. They are in detail: (3)  $^{12}$C(p,p$\gamma$), (4) 
$^{16}$O(p,p$\alpha$$\gamma$), (5) $^{12}$C($\alpha$,$\alpha$$\gamma$), (6)
$^{16}$O($\alpha$,2$\alpha$$\gamma$).}
\label{fig6}
\end{figure}

\begin{figure}
\epsfig{file=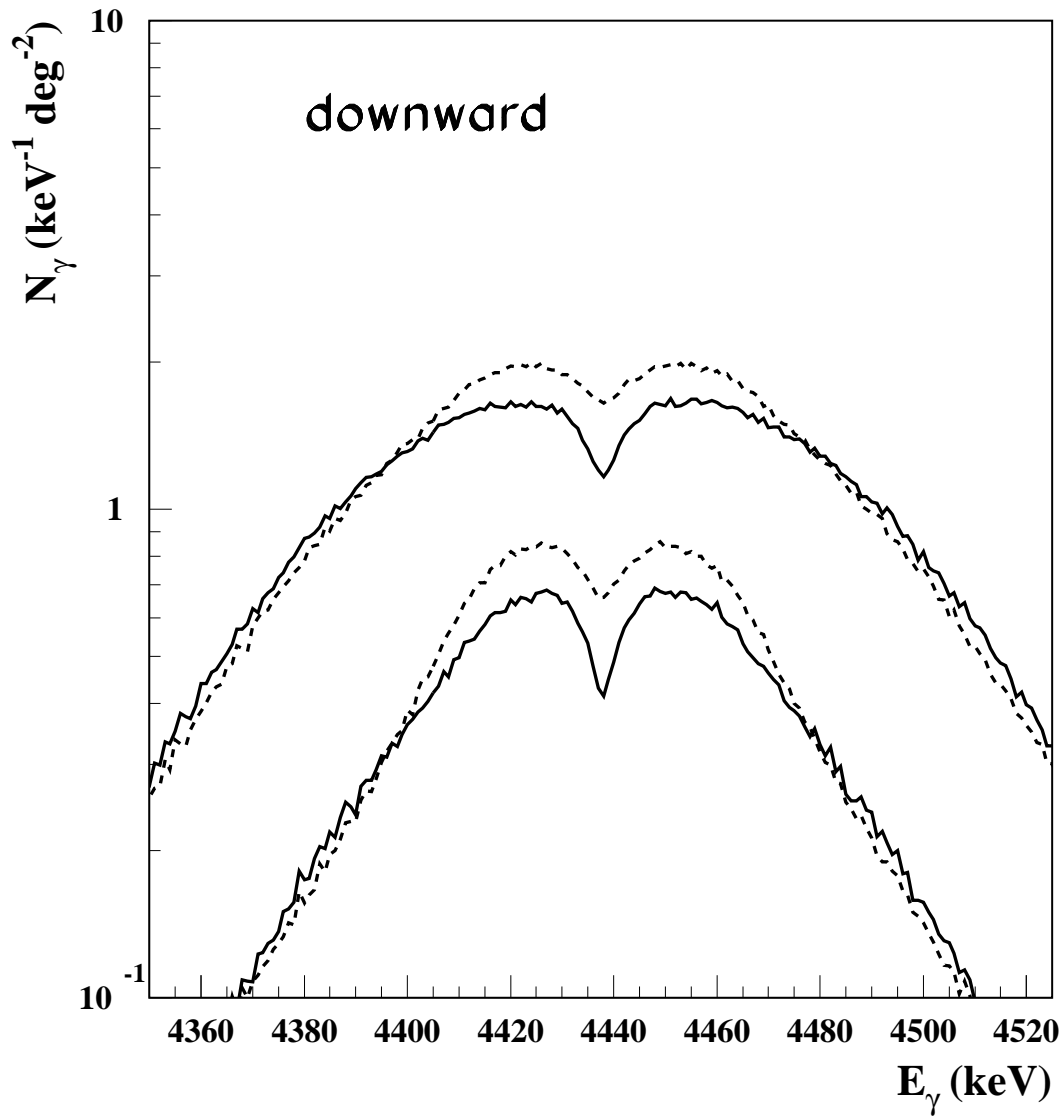}
\caption{
Emitted line profiles of the 4.438 MeV $\gamma$-ray for downward
flares at the solar limb, i.e. for a detection angle of 90$^{\circ}$ with
respect to the energetic particle direction. The upper two curves are for an
energetic particle spectrum with $s$=2.4; $E_0$=300 MeV, the lower two curves
are for $s$=3.3; $E_0$=30 MeV. Full lines are the profiles calculated with
population amplitudes from this work; for comparison, the dashed lines are
calculated with population amplitudes from the parameterization proposed by
Murphy, Kozlovsky and Ramaty [5]}

\label{fig7}
\end{figure}

\begin{figure}
\epsfig{file=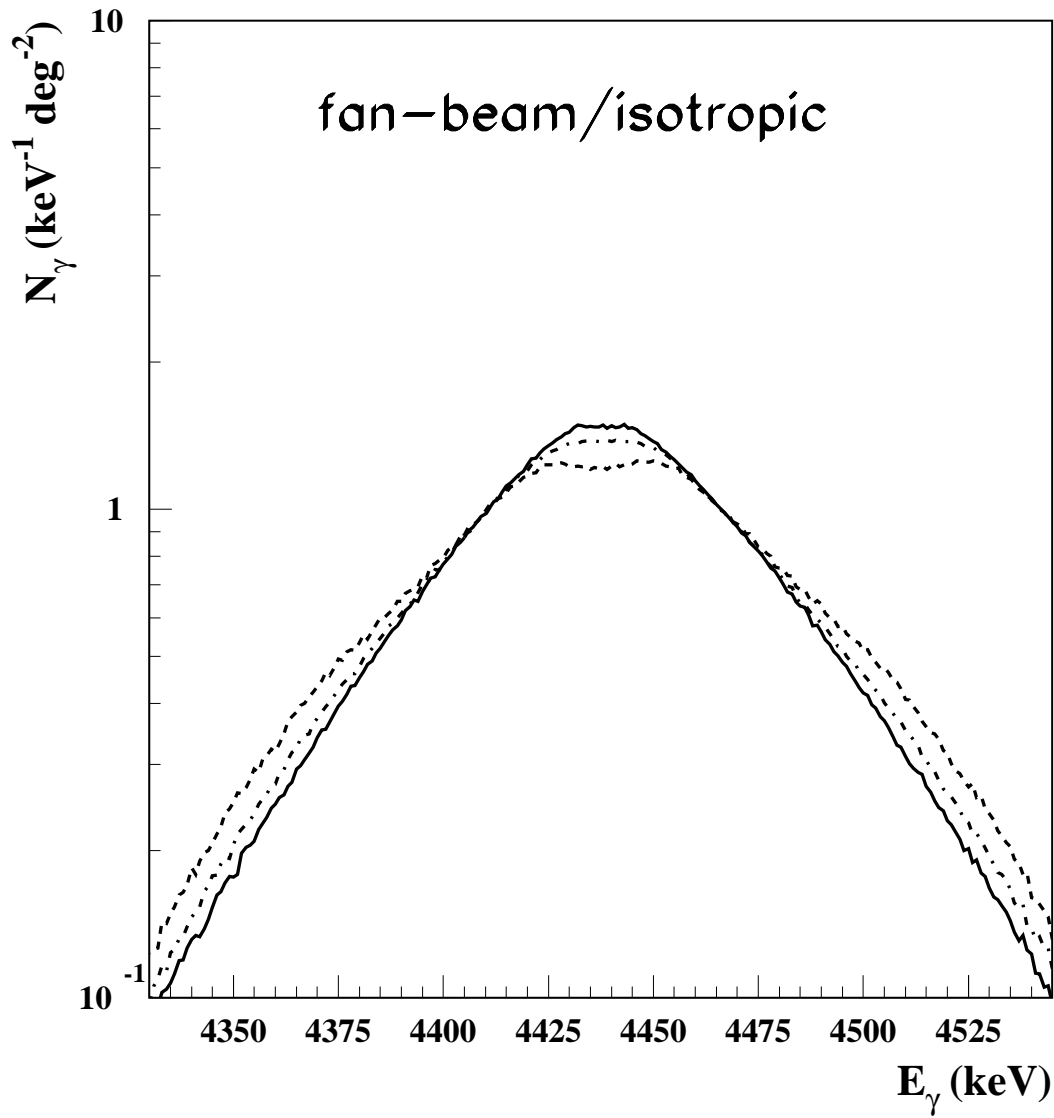}
\caption{
Emitted line profiles of the 4.438 MeV $\gamma$-ray for two
fan-beam flares and an isotropic flare with energetic particle spectrum
parameters $s$=3.3; $E_0$=30 MeV. The full line is for a fan-beam flare
occuring at the solar disk, with an angle $\Theta_{flare}$
between the line-of-sight and the flare symmetry axis of 135$^{\circ}$ and
an angular distribution proportional to sin$^6$($\Theta_p$),
$\Theta_p$ being the angle between the particle direction and the flare 
symmetry
axis. The dashed line shows the profile for the same flare occuring at the
solar limb ($\Theta_{flare}$=90$^{\circ}$), and the dashed-dotted line the
profile for an isotropic interaction probability.}
\label{fig8}
\end{figure}

It is the first parameterization for line shape calculations, which is
largely based on experimental data and optical model calculations.
Especially, the calculation of $\gamma$-ray line shapes from proton induced
reactions is certainly very close to reality, as illustrated by the good
reproductions of experimental line shapes in Figs. \ref{fig1} and \ref{fig4}; 
at higher
energies, the optical potential of Ref. \cite{Mei} is believed to provide line
shape reproductions of the same quality. Only line shapes from
$\alpha$-particle induced inelastic scattering below $\approx$ 10 MeV per
nucleon have to be regarded with some caution because of the only moderately
successfull OM fits of differential inelastic cross sections. Here, new
experimental data for $\gamma$-ray line shapes and angular distributions
could greatly improve the parameterization. For $\alpha$-particle induced
spallation of $^{16}$O above $E_{\alpha}$ = 26 MeV, $\gamma$-ray production
cross sections are needed.

This study confirms qualitatively older parameterizations for proton induced
reactions, constructed with much less experimental data input \cite{RKL}
\cite{MKR}, but some of its predictions of line-profile details are slightly
different. It shows, that extreme particle angular distributions, like
isotropic or unidirectional ones can be nicely characterized, with even some
hope to get an idea of the spectral hardness of the accelerated particle
spectrum. High resolution detectors, like the Ge detectors on the HESSI
spacecraft \cite{Hessi} and on the INTEGRAL observatory \cite{Int}, could be
able to resolve the fine structures in the line profiles which carry this
information. Together with other $\gamma$-ray lines, one may then be able
to extract most flare parameters, like the spectrum, composition and the
angular distribution of the accelerated particles, as well as the isotopic
composition of the ambient medium.

\newpage

\begin{table}
\caption{
Cross section $\sigma$ and mean excitation energies for the
spallation reaction
$^{16}$O(p,p$\alpha$)$^{12}$C$^{\star}_{4.439}$ and parameter $p$ of the
inelastic scattering angular distributions $^{16}$O(p,p').
\newline
($E_p$ = 16 - 20 MeV):
Mean excitation energies $\mathcal{E}$$_x$ in $^{16}$O
from the adjustment of the 4.438 MeV $\gamma$-ray  lines of the Orsay
experiment [8]. Isotropic
inelastic scattering angular distributions were used ($p$=0). 
\newline
($E_p$ = 22.5 - 100 MeV):
Mean excitation energies $\mathcal{E}$$_x$ in $^{16}$O  and
fit-parameter $p$ of the inelastic scattering angular distributions
$^{16}$O(p,p') obtained by OM calculations.
For $E_p$$<$16 MeV, the values of 16 MeV were used.}
\label{tab1}
\begin{tabular}  {lrll}
{$E_p$ (MeV)} & {$\sigma$ (mb)} & {$\mathcal{E}$$_x$ (MeV)} &
{$p$ (deg$^{-1}$)} \\ 
\hline
14 & 3 & 12.75 &.0 \\
15 & 16 & 12.75 & .0 \\
16 & 40 & 12.75 & .0 \\
17 & 64 & 13.25 & .0 \\
18 & 85 & 14.0 & .0 \\
20 & 140 & 14.75 & .0 \\
22.5 & 156  & 14.7 & .001 \\
25 & 137 & 15.5 & .008 \\
30 & 95 & 17.1 & .009 \\
40 & 55 & 18.0 & .027 \\
45 & 44 & 18.3 & .030 \\
66 & 24 & 18.5 & .057 \\
100 & 14 & 18.5 & .078 \\
\end{tabular}
\end{table}

\begin{table}
\caption{
Cross section $\sigma$ and population amplitudes for the inelastic
scattering reaction $^{12}$C(p,p)$^{12}$C$^{\star}_{4.439}$.
\newline
($E_p$ = 8.6 - 19.75 MeV):
Result of the adjustment of the population amplitudes $a_m$ and
the phase shift $\Delta\Phi$ to measured line shapes from the Orsay
experiment [8] and $\gamma$-ray angular
distributions from the Orsay experiment and from the Washington experiment
[11].
\newline
($E_p$ = 25 - 100 MeV): Result of the adjustment of the population 
amplitudes $a_m$ and
the phase shift $\Delta\Phi$ to line shapes and $\gamma$-ray angular
distributions resulting from OM calculations.
Population amplitudes below 8.6 MeV are an average of the extracted values
from 8.6-10.6 MeV.
\newline
The amplitudes are normalized such that $\sum_{m}  a_m^2  =  1$
and $a_{-m} = a_m$.  }
\label{tab2}
\begin{tabular}  {lrllll}
{$E_p$ (MeV)} & $\sigma$ (mb) &
$a_0$ & $a_1$ & $a_2$ & $\Delta\Phi$ (deg)
\\ \hline
5 & 51 & .64 & .27 & .47 & 30  \\
6 & 55 & .64 & .27 & .47 & 30  \\
7 & 108 & .64 & .27 & .47 & 30  \\
8 & 240 & .64 & .27 & .47 & 30  \\
8.6 & 270 & .57 & .28 & .51 & 35   \\
9.0 & 290 & .66 & .26 & .46 & 30  \\
9.6 & 275 & .71 & .35 & .35 & 30   \\
10.0 & 265 & .70 & .28 & .42 & 30  \\
10.6 & 296 & .59 & .21 & .53 & 20   \\
11.0 & 317 & .55 & .33 & .49 & 35  \\
11.4 & 298 & .56 & .34 & .45 & 30   \\
12.0 & 270 & .55 & .22 & .55 & 25  \\
12.6 & 277 & .42 & .08 & .63 & 25   \\
13.0 & 282 & .40 & .12 & .64 & 25  \\
13.6 & 281 & .42 & .13 & .63 & 25   \\
14.0 & 281 & .42 & .13 & .63 & 25  \\
14.4 & 255 & .42 & .13 & .63 & 25   \\
15.2 & 217 & .42 & .13 & .63 & 25  \\
16.25 & 188 & .33 & .03 & .66 & 20   \\
17.25 & 166 & .27 & .03 & .68 & 20  \\
18.25 & 153 & .27 & .16 & .66 & 15   \\
19.75 & 130 & .29 & .20 & .64 & 15  \\
25 & 99 & .25 & .025 & .68 & 112.5  \\
35 & 69 & .37 & .0 & .66 & 100   \\
49.5 & 23 & .43 & .0 & .64 & 100  \\
75 & 14 & .51 & .0  & .61 & 95   \\
100 & 11 & .58 & .0 & .58 & 90  \\
\end{tabular}
\end{table}

\begin{table}
\caption{
Estimated cross section $\sigma$ of the spallation reaction,
fit-parameters $p$ and
$C_2$/$C_1$ of  the inelastic scattering angular distributions
$^{16}$O($\alpha$,$\alpha$') and mean excitation energy $\mathcal{E}$$_x$
of $^{16}$O deduced from OM calculations for the  spallation reaction
$^{16}$O($\alpha$,2$\alpha$)$^{12}$C$^{\star}_{4.439}$.}
\label{tab3}
\begin{tabular}  {lrlrc}
{$E_{\alpha}$ (MeV)} & {$\sigma$ (mb)} & {$p$ (deg$^{-1}$)} &
{$C_2$/$C_1$ } & {$\mathcal{E}$$_x$ (MeV)} \\ \hline 
18 & 0.22 & .020 & .12 & 12.8 \\
20 & 19 & .017 & .01 & 12.8 \\
22 & 44 & .012  & -.08 & 13.0 \\
24 & 102 & .009  & .06 & 13.8 \\
26 & 125 & .008  & .27 & 14.2 \\
28 & 124  & .005 & .51 & 14.5 \\
30 & 123  & .003 & .61 & 14.8 \\
35 & 120 & .001  & 1.91 & 15.7 \\
40 & 117 & .010  & .94 & 16.3 \\
50 & 111 & .019  & .99 & 16.6 \\
60 & 139 & .015  & .42 & 17.1 \\
100 & 223 & .074 & .0 & 18.4 \\
150 & 172 & .125 & .0 & 18.8 \\
250 & 100 & .31  & .0 & 19.0 \\
400 & 53  & .51  & .0 & 19.0 \\
\end{tabular}
\end{table}

\begin{table}
\caption{
Cross section $\sigma$ and 
results of the adjustment of the population amplitudes $a_m$ and
the phase shift $\Delta\Phi$ to measured and calculated line shapes and
$\gamma$-ray angular distributions
for inelastic $\alpha$-particle scattering to the 4.439 MeV, 2$^+$ state of
$^{12}$C. The amplitudes are normalized such that $\sum_{m}  a_m^2  =  1$
and $a_{-m}=a_m$; $a_{\pm1}$=0. }
\label{tab4}
\begin{tabular}  {lrllll}
{$E_{\alpha}$ (MeV)} & $\sigma$ (mb) &
$a_0$ & $a_1$ & $a_2$  & $\Delta\Phi$ (deg) \\ \hline
8 & 48 & .0 & .0 &  .71 & -5   \\
9 & 72 & .0 & .0 & .71 & -5  \\
10 & 231 & .0 & .0 & .71 & -5   \\
11 & 347 & .0 & .0 & .71 & -5  \\
12 & 393 & .0 & .0 & .71 & -5   \\
13 & 361 & .0 & .0 & .71 & -10  \\
14 & 431 & .14 & .0 & .70 & -22.5   \\
15 & 381 & .0 & .0 & .71 & -12.5  \\
16 & 371 & .17 & .0 & .70 & -22.5   \\
17 & 306 & .27 & .0 & .68 & -32.5  \\
21 & 276 & .68 & .0 & .52 & -102.5   \\
25 & 292 & .65 & .0 & .54 & -90  \\
32.5 & 141 & .61 & .0 & .56 & -90   \\
41 & 94 & .37 & .0 & .66 & -55  \\
90 & 45 & .37 & .0 & .66 & -70   \\
139 & 27 & .45 & .0 & .63 & -70  \\
166 & 26 & .40 & .0 & .65 & -75   \\
250 & 18 & .37 & .0 & .66 & -80  \\
400 & 6 & .58 & .0 & .58 & -80  \\
\end{tabular}
\end{table}

\newpage 

\end{document}